\documentclass[12pt]{iopart}
\usepackage{bm}
\usepackage{graphicx}

\usepackage{iopams} 
\usepackage{cite}

\begin{document}
\title[Landau damping of Alfv\'enic modes in stellarators]{Landau damping of Alfv\'enic modes in stellarators}
\author{Ya.I. Kolesnichenko, A.V. Tykhyy}
\address{Institute for Nuclear Research, Prospekt Nauky 47, Kyiv 03028, Ukraine}

\begin{abstract}
It is found that the presence of the so-called non-axisymmetric resonances of wave-particle interaction in stellarators
[which are associated with the lack of axial symmetry of the magnetic configuration,
\textit{Kolesnichenko et al.,  Phys. Plasmas \textbf{9} (2002) 517}]
may have a strong stabilizing influence through Landau mechanism
on the Toroidicity-induced Alfv\'en Eigenmodes (TAE) and isomon modes
(Alfv\'enic modes with equal poloidal and toroidal mode numbers and frequencies in the continuum region)
destabilized by the energetic ions. These resonances involve largest harmonics of the equilibrium magnetic
field of stellarators and lead to absorption of the mode energy by thermal ions in medium pressure plasma,
in which case the effect is large. On the other hand, at the high pressure attributed to, e.g., a Helias reactor,
thermal ions can interact also with high frequency Alfv\'en gap modes
[Helicity-induced Alfv\'en Eigenmodes (HAE) and mirror-induced Alfv\'en Eigenmodes (MAE)],
leading to a considerable damping of these modes.
Only resonances with passing particles are considered.
The developed theory is applied to various modes in the Wendelstein 7-X stellarator and
a Helias reactor, and to two TAE modes in the LHD helical device.
\end{abstract}
\submitto{\PPCF}
\date{\today}
\section{Introduction}
\label{sec:intro}

Effects of the energetic ions on plasma stability are determined by the competition of  the energy transfer  from these ions to the waves  and absorption of the wave energy by the bulk plasma particles. Several damping mechanisms can lead to the absorption of the wave energy. These mechanisms are Landau damping, continuum damping, radiative damping, and collisional damping.  A brief overview of them in stellarators can be found in, e.g.,  \cite{PPCF}, see also a more recent work \cite{Slaby}. The physics of the  damping is rather similar   in both tokamaks and stellarators.  However, there are important differences associated with different structures of the magnetic field in the mentioned toroidal systems: First, the presence of many Fourier harmonics in the magnetic field strength and the variation of the plasma cross section shape along the large azimuth of the torus in stellarators increase the number of gaps in  Alfv\'en continuum, leading  to more types of eigenmodes 
  \cite{Naka,Nuehrenberg,AE}, see also the overview \cite{PPCF}.  Namely, in addition to Toroidicity-induced 
	Alfv\'en eigenmodes (TAE), 
Ellipticity-induced  Alfv\'en Eigenmodes (EAE), and  noncircular triangularity-induced Alfv\'en Eigenmodes (NAE) existing in tokamaks, in stellarators there are Mirror-induced Afv\'en eigenmodes and various Helicity-induced 
 Afv\'en eigenmodes (HAE).   Second,  in stellarators  there exist so-called non-axisymmetric resonances which provide an additional  wave-particle interaction \cite{PPCF,AI}. Modes with frequencies in the continuum region, known as Global  Afv\'en eigenmodes (GAE) and  Non-conventional  Global  Afv\'en eigenmodes (NGAE) can also be excited \cite{GAE1,GAE2,NGAE}. 
The presence of transitioning particles considerably affects the collisional damping of Alfv\'en eigenmodes \cite{March}.
Recently it was found that low-shear stellarators with the rotation transform close to unity, such as Wendelstein 7-X,   are    prone to Isomon Modes (IM), which are Alfv\'enic modes affected by plasma compressibility and having equal poloidal and toroidal mode numbers, $m=n$ \cite{isomon}. The radial width of the IM modes is rather large, these modes extend over a large part of the plasma cross section. Their  destabilization in W7-X by passing energetic ions (with the maximum energy $55$ - $60$ keV) produced as a result of the Neutral Bream Injection (NBI) was considered in \cite{isomon}. It was concluded that the IM instability  growth rate in the first planned NBI experiments on W7-X can be rather large. However, the only wave-particle interaction taken into account was the resonance interaction  between the modes and NBI ions, the interaction with the bulk plasma particles which may lead to absorption of the mode energy was ignored. No other damping mechanisms was considered. Therefore, it was not clear whether the NBI  will really lead to the isomon instability. 

	The purpose of this work is to consider Landau damping of the IM modes and Alfv\'en gap modes in stellarators, first of all,  in Wendelstein 7-X. An important role of this damping mechanism in stellarators was recognized in \cite{PPCF}: It was drawn attention to the fact that, as shown in the earlier work \cite{AI}, non-axisymmetric resonances can lead to rather low resonance velocities and, therefore,  thermal particles can absorb the wave energy. Till now, however, there were no comprehensive studies of Landau damping in stellarators, although some steps in this direction were already done \cite{Slaby,Koenies}.   

The structure of the work is the following. Resonances of the wave-particle interaction in stellarators are considered in section \ref{sec:quality}. In this section the analysis is based on the equation describing resonances of passing particles and the waves \cite{AI}, which was applied to Alfv\'en gap modes and IM modes in Wendelstein 7-X. In  section \ref{sec:gr} general expressions for the instability growth / damping rate of Alfv\'enic modes, including the case of modes  in compressible plasmas, are derived. They are applicable to study effects  of both electrons and ions (superthermal and thermal) on the modes.  In section \ref{sec:examples} the derived equations are used to calculate the damping rates and growth rates of the IM modes and Alfv\'en gap modes,  the calculations are carried out  both in the local approximation and with  relations taking into account the radial structure of the modes. All the specific examples are relevant to Wendelstein 7-X and a Helias reactor, except for a one relevant to the LHD device. The obtained results are summarized in section~\ref{sec:sum}.

\section{Analysis of  resonances between  Alfv\'enic modes and passing particles}  
\label{sec:quality}

\subsection{Resonance equation}
\label{subsec:res}

 Resonance interaction between Alfv\'enic modes  and the particles can lead not only to destabilization of these modes but it plays an important role in the mode damping, in particular, due to Landau mechanism. This mechanism is especially important for stellarators, as will be shown below.

 We employ the following Fourier serious for the magnetic field (B),  the field line curvature  ($\bcal{K}$), and a perturbed quantity labeled by tilde \cite{AE,PPCF,AI}: 
\begin{equation}
 B=\bar{B} \left (1+\frac{1}{2} \sum_{\mu,\nu}\epsilon_{\mu\nu}e^{i\mu\vartheta -i\nu N\varphi}  \right ),
   \label{B0}
\end{equation}
 \begin{equation}
 \bcal{K}=\sum_{\mu,\nu} \bcal{K}_{\mu,\nu}(r) e^{i\mu \vartheta -iN\nu\varphi} ,
   \label{K}
\end{equation}
\begin{equation}
 \tilde{X}=\sum_{m,n} X_{m,n}(r)e^{im\vartheta -in\varphi -i\omega t},
   \label{X}
\end{equation}
 where $\bar{B}$ is the average magnetic field at the magnetic axis, the radial coordinate $r$ is defined by $\psi = \bar{B}r^2/2$, $\psi$ is the toroidal magnetic flux,  $\vartheta$ and $\varphi$ are the poloidal and toroidal Boozer angles, respectively,   $N$ is the number of periods of the equilibrium field.

Using these notations, the resonance between the waves and most passing particles in stellarators can be described as follows \cite{AI}:  
\begin{equation}
\omega = k_{res} \,v_\parallel^{res},
\label{res1}
\end{equation}
where    $v_\parallel^{res} $ is the particle resonance velocity along the magnetic field, 
$k_{res}\equiv k_{m+\mu, n+\nu N} = [(m+\mu)\iota -(n+\nu N)]/R$, $\iota$ is the rotational transform of the field lines, $\iota =1/q$, $q$ is the tokamak safety factor,  
and $R$ is the major radius of the torus.  Equation  (\ref{res1}) was obtained within a theory of destabilization of Alfv\'en eigenmodes by fast ions.   It can also be  derived  by proceeding from the equations 
\begin{equation}
\frac{d\mathcal{E}}{dt} = e\mathbf{v}_D\cdot\tilde{\mathbf{E}},
\label{initial}
\end{equation}
\begin{equation}
\vartheta (t) =\omega_\vartheta t +\vartheta_0,\;\;\; \varphi (t) =\omega_\varphi t +\varphi_0,
\label{motion}
\end{equation}
where  $\mathbf{v}_D$ is the particle drift velocity in the stellarator magnetic field,  
$ \tilde{\mathbf{E}}=\hat\mathbf{E}_\perp(r)\exp (-i\omega t +im\vartheta -in\varphi)$ is the electric field of a wave; $\omega_\vartheta$  and $\omega_\varphi$ are the frequencies of the particle motion in the poloidal and toroidal directions, respectively \cite{PPCF}. 
 Equation (\ref{initial}) describes the energy exchange between a wave with the electric field $ \tilde{\mathbf{E}}$
 and a charged particle moving across the magnetic field with the drift velocity $\mathbf{v}_D$.  Equation (\ref{motion}) describes the particle motion along the magnetic field.

For given mode frequency and mode numbers,  equation (\ref{res1}) determines the infinite number of the  resonance  velocities 
$v^{res}_\parallel $  for which  particles  can interact with the modes. However, only several of them are important, depending on  the number of considerable Fourier harmonics of the magnetic field.  Note that the mirror harmonic of the magnetic field has a minor influence on the damping, although it is the largest harmonic in the plasma core in Wendelstein 7-X (especially, in the high mirror configuration).  This conclusion follows from the  relations for the damping rate which will be derived in the work.

\subsection{Resonances between particles and Alfv\'en gap modes}
\label{subsec:rgm}

In this subsection, we analyze resonances for eigenmodes with frequencies located inside the gaps in Alfv\'en continuum (AC).

 In the presence of a Fourier harmonic with the numbers   $(\mu_0,\nu_0)$ in the magnetic field and / or in the metric tensor, a gap in the Alfv\'en continuum arises at the radius $r_*$ where two cylindrical branches  with the mode numbers  $m,\,n$ and $m+\mu_0,\,n+\nu_0$ 
[$\omega_1= |k_{mn}|v_A (r)$ and $\omega_2 =|k_{m+\mu_0 ,n+\nu_0 N}|v_A(r)$, $v_A$ is the Alfv\'en frequency] intersect. This takes place at \cite{AE}
\begin{equation}
\iota_* = \frac{2n+\nu_0 N}{2m +\mu_0}.
\label{gapiota}
\end{equation} 
The numbers  $\mu_0$ and $\nu_0$ label Fourier harmonics relevant to Alfv\'en continum and Alfv\'en eigenmodes, in contrast to the 
 $(\mu$, $\nu)$ numbers relevant the resonance.    
 The frequency of the considered branches at $\iota=\iota_*$ can be written as  
\begin{equation}
\hat{\omega} = |k_{mn}^{\omega}|v_{A*}= 0.5|k_{\mu_0\nu_0}^{\omega}|v_{A*},
\label{gapfreq}
\end{equation}
where $k_{mn}^{\omega}=-0.5k_{\mu_0\nu_0}^{\omega}$,   $k_{mn}^{\omega}=(m \iota_* -n)/R_{\omega}$,  $k_{\mu_0\nu_0}^{\omega}=(\mu_0 \iota_* -\nu_0 N)/R_\omega$, $v_{A*}=v_A(\iota_*)$, and $R_\omega =R $ at the intersection point in the space $(r,\omega)$. With this notations, the resonance  (\ref{res1}) at $\iota =\iota_*$  takes the form:
 $\omega = [-0.5 k_{\mu_0\nu_0} +k_{\mu\nu}]v_\|^{res}$, where  $k_{\mu\nu} = k_{\mu\nu}^{\omega}$ with   $R_\omega =R$.
Note that equation (\ref{gapfreq}) was obtained without involving any specific magnitude of the major radius of the torus.  Due to this,   the radius $R_\omega $  can be  considered as an adjustable parameter allowing to describe by  (\ref{gapfreq}) not only the continuum frequency of two cylindrical branches but also  a mode frequency in realistic magnetic configurations.

The frequency of a mode which can reside in the gap, in general, does not equal to $\hat{\omega}$ but is close to it when the gap  is narrow. In stellarators, however, there are wide gaps and, moreover, the gaps  can be considerably shifted. In particular, the TAE gap in the AC of the Wendelstein-line stellarators is shifted down  because of the interaction with other gaps (first of all, with the very wide gap associated with  the  $\mu_0 /\nu_0 =2/1 $ helical shaping of the plasma cross section, see, e.g., figure 2 in \cite{PPCF}) located above. This effect can be taken into account  by assuming  $R_\omega >R$. On the other hand, the plasma compressibility (which produces the low frequency $\beta$-induced gap) tends to shift  gaps in the AC up. This effect seems weak in stellarators but can be very strong in spherical tokamaks and in conventional tokamaks with hollow current when 
 $\beta >\iota^2$ ($\beta =8\pi p /B^2$, $p$ is the plasma pressure) \cite{Fesen}.

Combining equation (\ref{res1}) and  equation (\ref{gapfreq}) we obtain: 
\begin{equation}
v_\|^{res} = \frac{R}{R_\omega} \left (\mbox{sgn}\, k_{mn} +2\frac{\mu\iota_* -\nu N}{|\mu_0 \iota_* -\nu_0 N|}\right )^{-1} v_{A*}.
\label{vres}
\end{equation}
We observe that $|v^{res}_{\|}| =v_{A*}$ and  $|v^{res}_{\|}| =v_{A*}/3$ for $\mu =\mu_0$, $\nu =\nu_0$ and $R=R_\omega$  at the radius where $\iota =\iota_*$. 
This implies that the  resonance velocities are the same for all the gap modes, provided that corresponding Fourier harmonics of the magnetic field are present
(we remind that some gaps in the AC are produced by the plasma shaping rather than by harmonics of the magnetic field). 
On the other hand, when  $\mu \neq\mu_0$ or $\nu \neq \nu_0$, there are a variety of resonance velocities even for $R=R_\omega$.

 Let us consider specific examples. 

First of all, we consider a  plasma with high $\beta$ (the ratio of the plasma pressure to the magnetic field pressure) required for a good confinement of the energetic ions. In particular,  $\beta (0) =13 -14 \%$ is expected in a Helias reactor  \cite{Wobig}.  Taking  for the bulk plasma ions 
$\beta_i\equiv 8\pi n_iT_i /B^2 =6.5\%$ we obtain that the mentioned above resonance velocity, $v_\parallel^{res} =v_A /3$,  provides an efficient interaction of various Alfv\'en  gap modes and the bulk plasma ions because in this case the resonance velocity is close to the thermal velocity of the ions, 
$v_\|^{res} /v_{Ti} =1.3$.  

Below, however,  we restrict ourselves to plasmas relevant to current experiments and  plasmas expected at the initial stage of operation with NBI on  Wendelstein 7-X.  We assume that in the region of the mode location  
\begin{equation}
\beta_i \sim \frac{1}{4N^2},
\label{beta}
\end{equation}
 in which case $v_\|^{res} \sim v_{Ti}$, where  $v_{Ti} =\sqrt{2 T_i/M_i}$ is thermal velocity  of the bulk plasma ions. 
 This estimate follows from (\ref{vres}) for a TAE mode interacting with the bulk plasma ions through the helical resonance  with $\mu /\nu =1/1$   when $N\gg 1$.
 We refer to plasmas satisfying (\ref{beta})  as a low-beta case.

\begin{figure}
\centering
\includegraphics[width=8cm]{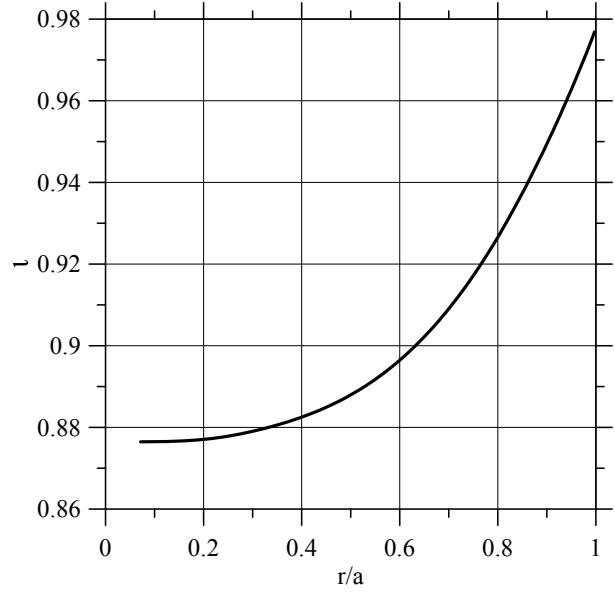}
\caption{The rotational transform of the field lines for $\beta(0)=0.037$ in the W7-X 
high-mirror configuration.}
\label{Fig:iota}
\end{figure}

 In particular, in Wendelstein 7-X  where $N=5$ and $\iota_* \approx 0.9$ (see figure 
\ref{Fig:iota}),  the $\mu=\nu =1$ resonance at $\iota =\iota_*$ leads to $(R_\omega /R)|v^{res}_{\|}| /v_A{*} = 1/10.1$ and  $1/8.1$. Hence,
 \begin{equation}
\frac{|v^{res}_{\|}|}{v_{Ti}} = \frac{R}{ 10R_\omega \sqrt{\beta_i}}.
\label{taehae}
\end{equation}
As expected,  this resonance velocity is close to thermal velocity  of the bulk plasma ions 
 for  $\beta_i =0.01$. On the other hand, the helical harmonic with $\mu =\nu =1$ is rather large.  Therefore, one can expect that the helicity-induced resonance will have a strong stabilizing influence on  the TAE instability in Wendelstein 7-X. 

 High frequency modes (HAE$_{11}$, HAE$_{21}$ and MAE modes) can be damped due to the tokamak sideband resonance, $\mu /\nu =1/0$. Using  the same rotational transform, $N=5$, and assuming  $R=R_\omega$,  we obtain $|v_\|^{res}|/v_{A*} =0.735$ and $1.56$ for MAE,  $|v_\|^{res}|/v_{A*} =0.69$ and $1.78$  for HAE$_{11}$,  $|v_\|^{res}|/v_{A*} =0.64$ and $2.28$  for HAE$_{21}$. All these magnitudes of the resonance velocity well exceed the ion thermal velocity,  which implies that the ion damping will be exponentially small.    
In contrast, because the electron thermal velocity, $v_{Te}$, typically exceeds Alfv\'en velocity,  the electron damping can be considerable:
 \begin{equation}
\frac{|v^{res}_{\|}|}{v_{Te}} = \frac{|v^{res}_\||}{v_{A*}}\sqrt\frac{M_en_{i*}}{M_in_{e*}} \frac{1}{ \sqrt{\beta_e}}\frac{R}{R_\omega},
\label{haetae}
\end{equation}
where  $\beta_e = 8\pi n_{e*}T_{e*} /B^2$. 

In a hydrogen plasma, the magnitudes of the ratio  $R_\omega|v_\|^{res}|/(Rv_{Te})$ are the following:  $0.17$ and $0.36$ for MAE,  $0.16$ and $0.41$  for HAE$_{11}$,  $0.14$ and $0.52$  for HAE$_{21}$. We conclude that the electron damping of these modes will be  not so strong as the ion  damping of     TAE modes caused by the $(\mu /\nu=1/1)$-helical resonance. The matter is that, first, the ratio 
$|v_\|^{res}/v_{Te}|$  for HAEs and MAEs is much less than unity (unless $R>R_\omega$), whereas  for TAEs  $|v_\|^{res}/v_{Ti}|\sim 1$ and, second, the toroidal Fourier harmonic, $\epsilon_{10}$, in W7-X is less than the helical harmonic, $\epsilon_{11}$, by a factor of two  (as will be shown below, the damping rate is proportional to $\epsilon_{\mu\nu}^2$).

Considered examples are relevant to a hydrogen plasma. In plasmas with more heavy ions the ratio  $|v_\|^{res}|/v_{Te}$ is smaller, as follows from equation (\ref{haetae}).  Therefore, the  electron damping  is smaller, too.  

Damping of the MAE mode in W7-X is affected also by the helical harmonic $\epsilon_{11}$.

\subsection{Resonance between particles and isomon modes}
\label{subsec:rim}

It follows from  equation (\ref{res1}) that due to a helical resonance, the IM modes  interact with the particles having the longitudinal velocity  given by 
\begin{equation}
v_\parallel^{res} =\frac{ R\omega} {-m\Delta\iota + \mu\iota -\nu N},
\label{vr}
\end{equation}
where $\Delta \iota =1-\iota$. Frequencies of  the IM modes slightly exceed the frequency $\omega= m|\Delta \iota |v_{Am}/R$ (at least, for $m>1$), with $v_{Am}$ the Alfv\'en velocity at the radius where the mode amplitude is maximum. Hence, the ratio of the ion resonance velocity to  thermal velocity is  
\begin{equation}
\left |\frac{v_\parallel^{res}}{v_{Ti}}\right | =\frac{ |m|\Delta \iota} {|-m\Delta\iota + \mu\iota -\nu N|\sqrt{\beta_i}}.
\label{rt}
\end{equation}

 Taking and $\Delta \iota =0.1$ and $\beta_i =0.01$ we obtain $v_\|^{res}/v_{Ti} \lesssim 1$ from equation  (\ref{rt}) for $m=2-4$. 
The frequency of the $m=1$ mode exceeds the magnitude $m\Delta \iota v_A /R$ by a factor of $2.4$ (because of plasma compressibility) \cite{isomon}.   Due to this,  $v_\|^{res}$ for the m=1 mode is also close to the ion thermal velocity. Therefore, one can expect that the ion damping of isomon modes will be considerable.

\section{Derivation of relations for the mode damping / growth  rate}
\label{sec:gr}

\subsection{Relations for  Alfv\'en modes}
\label{subsec:1}

In this subsection we derive a general expression for the damping / growth rate of Alfv\'en eigenmodes in a plasma containing energetic ions by using
 a perturbative approach.

We proceed from the following quasi-neutrality  equation: 
\begin{equation}
\nabla \cdot \tilde\mathbf{j} =0,
\label{nj}
\end{equation}
 where
 $\mathbf{j}$ is the plasma current, tilde above a letter here and below labels perturbed quantities. Multiplying
 equation ($\ref{nj}$) by a scalar potential of the electromagnetic field, $\tilde{\Phi}$, and integrating the product over the plasma volume we obtain
\begin{equation}
\int d^3x \tilde{\mathbf{j}}\cdot \nabla \tilde{\Phi} =0
\label{int}
\end{equation}
provided that $\tilde{\Phi}$ satisfies the equation $\int d\mathbf{s} \cdot \tilde{\mathbf{j}}\tilde{\Phi} =0$, where the integral is taken over at the plasma boundary. 

The  current  $\tilde{\mathbf{j}}$ in the integrand of (\ref{int}) is convenient to  write   as 
$\tilde{\mathbf{j}} = \tilde{\mathbf{j}}_\| +\tilde{\mathbf{j}}_\perp$, 
where the subscripts $\parallel$ and $\perp$ label  magnitudes along and across the magnetic field, respectively. Let us 
determine the longitudinal current from Maxwell equations for the electromagnetic field. Then the  transverse current should be determined from the plasma equations. Namely,  for  $\tilde{\mathbf{j}}_\|$ we will use the equations 
$c\nabla \times \tilde{\mathbf{B}} = 4\pi \tilde{\mathbf{j}}$ and $\tilde{\mathbf{B}} =\nabla \times \tilde{\mathbf{A}}$, with  $\mathbf{B}$ a magnetic field,
 $\mathbf{A}$ is the vector potential of the electromagnetic field. 
The transverse current in the presence of the energetic particles is  $\tilde{\mathbf{j}}_\perp = \tilde{\mathbf{j}}_\perp^{MHD} +  \tilde{\mathbf{j}}_\perp^{kin}+\tilde{\mathbf{j}}_\perp^\alpha$, where
  $\tilde{\mathbf{j}}_\perp^{MHD}$ is the plasma current in the framework of the ideal MHD,  $\tilde{\mathbf{j}}_\perp^{kin}$ is the kinetic part of the bulk plasma current, $\tilde{\mathbf{j}}^\alpha_\perp$ describes the fast ion current.
	
	In Alfv\'en waves  $\tilde{\mathbf{B}}_\|$ is small and, therefore,   $\tilde{\mathbf{A}}_\perp$ is small, too,  and can be neglected. 
Then 
\begin{equation}
\mathbf{\tilde{j}}_\| \approx -\frac{c }{4\pi  B_0}[\nabla \cdot B_0(\nabla_\perp\tilde{A}_\|  )]\mathbf{b},  
\label{jpar}
\end{equation}
where $\mathbf{b} =\mathbf{B}_0/B_0$ is the unit vector along the magnetic field, the subscript "0" labels equilibrium quantities.   
 On the other hand, ideal MHD equations with  $\mathbf{\tilde{E}} =-\nabla_\perp \tilde{\Phi}$  and $\tilde{\Phi} \propto \exp (-i\omega t)$ yield
\begin{equation}
\mathbf{\tilde{j}}_\perp^{MHD} \approx \frac{i\omega c^2}{4\pi v_A^2}\nabla_\perp \tilde{\Phi}.  
\label{jper}
\end{equation}

Only waves that are weakly damped in the absence of the energetic ions can be destabilized by a small group  of these  ions, so that a problem of stability can be treated perturbatively (the exception are  energetic particle modes, EPM, which are not considered 
here). This gives us grounds to use a perturbative approach in the analysis of equation (\ref{int}). Before doing it, we note that the longitudinal current, like the transverse  current, can be expressed in terms of the scalar potential $\tilde{\Phi}$. This can be  done by means of the relation  $\omega \tilde{A_\| } =ck_\|\tilde{\Phi}$ which follows from the ideal MHD equation $\tilde{E}_\| =0$ ($k_\parallel$ is defined by $ik_\| \tilde{\Phi}=\mathbf{b}\cdot \nabla \tilde{\Phi}$). Then $\tilde{j}_\parallel \propto \omega^{-1} $ and hence, 
$d(\omega \tilde{j}_\| )/d\omega =0 $.  Taking this into account, we write 
$\omega\tilde{\mathbf{j}}  = \omega\tilde{\mathbf{j}}^{(0)} + \omega \tilde{\mathbf{j}}^{(1)}$, with $\tilde{\mathbf{j}}^{(0)} =\tilde{\mathbf{j}}_\|+ \tilde{\mathbf{j}}_\perp^{MHD} $ and  
 $\tilde{\mathbf{j}}^{(1)} = \tilde{\mathbf{j}}^{kin}_\perp+\tilde{\mathbf{j}}^\alpha_\perp$,  $\tilde{\mathbf{j}}^{(1)}$ being small compared to $\tilde{\mathbf{j}}^{(0)}$. In zero approximation, we obtain from (\ref{int})  the following equation which is satisfied for the ideal MHD eigenfrequencies ($\omega_0$) and eigenmodes: 
\begin{equation}
\int d^3x \tilde{\mathbf{j}}^{(0)}(\omega_0) \cdot \nabla \tilde{\Phi} =0.
\label{eigen}
\end{equation}   
In the first approximation,  $(\omega\tilde{\mathbf{j}})^{(1)}  = [d(\omega_0\tilde{\mathbf{j}}_\perp^{(0)}) /d\omega_0 ]\Delta \omega  + 
\omega_0 \tilde{\mathbf{j}}^{(1)}$, with $\Delta \omega = \omega - \omega_0$. Defining the mode growth / damping rate as 
$\gamma =\mbox{Im} \,\Delta \omega $ and taking into account (\ref{jper}), we obtain in this approximation from equation (\ref{int}) after time averaging:
\begin{equation}
\gamma = \frac{0.5\mbox{Re}\int d^3x( \tilde{\mathbf{j}}_\perp^\alpha +  \tilde{\mathbf{j}}_\perp^{kin}) \cdot \nabla_\perp \tilde{\Phi}^*}
{\mbox{Re}\{(i\omega)^{-1}\int d^3x [\partial (\omega\tilde\mathbf{j}_\perp^{MHD})/\partial \omega]\cdot \nabla_\perp \tilde{\Phi}^* \}},
\label{gamma}
\end{equation}
where  the subscript $"0"$ at $\omega$ is omitted, the denominator equals to $2\mathcal{W_A}$, with $\mathcal{W_A}$ the Alfv\'en mode energy,
\begin{equation}
\mathcal{W_A} =\int d^3x \frac{c^2}{8\pi v_A^2} |\tilde{E}|^2.
\label{W}
\end{equation}
When obtaining these relations we took into account that Re $(\tilde{X}\tilde{Y}) =0.5$Re $(XY^*)$. 

Note that equation (\ref{gamma}) differs from the corresponding equation in reference \cite{AI} (equation (7) in \cite{AI}): The denominator  in (\ref{gamma})        contains derivative 
$d(\omega \tilde{\mathbf{j}}_\perp^{MHD})/ (d\omega )$ instead of  $d(\tilde{\mathbf{j}}^{MHD})/ (d\omega )$; this form of the mode energy is preferable because it does not involve the longitudinal current $\tilde{j}^{MHD}_\|$  and $\nabla_\| \tilde{\Phi}$. In addition, 
the numerator of (\ref{gamma}) contains the kinetic part of the bulk plasma current.

\subsection{Relations for modes in compressible  plasmas}
\label{subsec:2}

The quasi-neutrality equation used in subsection \ref{subsec:1} and the accuracy of equations (\ref{jpar}), (\ref{jper})  are not sufficient for the description of isomon modes in quasi-isodynamic stellarators, in particular, in  Wendelstein 7-X   and a Helias reactor \cite{Wobig}. These modes are determined by  equations for  potential $\tilde{\Phi}$ and compressibility $\tilde{\zeta}$  ($\tilde{\zeta} =\nabla \cdot \bm{\mathcal{\xi}}$, with 
$\bm{\mathcal{\xi}}$ the plasma displacement)   coupled  due to the field line curvature and finite plasma temperature \cite{isomon}. Therefore, in this subsection we derive a relation similar to (\ref{gamma}) by proceeding from the equation for isomode modes of \cite{isomon}  supplemented with a kinetic term associated with the bulk plasma (to be able to calculate the mode damping / growth rate).

This equation has the form:
  \begin{eqnarray}
 {1\over r}{d\over dr}r\delta_0\left({{\omega^2-\omega_{G}^2}\over v_A^2} -k_{mn}^2 \right ){d\Phi_{m,n}\over dr} \nonumber \\
-\left[{m^2\delta_0\over r^2} \left({{\omega^2-\omega_{G}^2r^2\epsilon_t^{\prime 2}/\epsilon_t^2}\over v_A^2} -k_{mn}^2 \right ) +
{k_{mn}\over r} (r\delta_0k_{mn}^\prime)^\prime \right ]\Phi_{m,n} \nonumber \\
 -\frac{4\pi i\omega }{c^2}B^r_{mn}\frac{d}{dr}\frac {j_{0\parallel}}{B} =\frac{4\pi i\omega}{c^2}[\nabla\cdot 
(\tilde{\mathbf{j}}_\perp^{\alpha}+\tilde{\mathbf{j}}_\perp^{kin}]_{ m,n},
  \label{eq1}
 \end{eqnarray}
where
\begin{equation}
 \omega_{G}^2  = \tilde{\epsilon}^2\frac{c_s^2}{R^2}\sum_{l=\pm 1} \left (\frac{\omega^2}{\omega^2 -k_{m+l,n}^2c_s^2}\right ),
  \label{G1}
 \end{equation}
 $\Phi_{m,n}$ is a Fourier component of $\tilde{\Phi}$,   
 $k_{mn} \equiv k_\parallel (m,n)=(m\iota -n)/R $,  $c_s =\sqrt{\Gamma p/\rho}$ is the sound velocity ($\Gamma =5/3$ is the heat capacity ratio, $p$ is the plasma pressure, and $\rho$ is the plasma mass density),
 $\delta_0\gtrsim 1$ is determined by the plasma shaping (see Ref. \cite{AE}), $\tilde{\epsilon}^2 =\epsilon_t^2/(\delta_0 \epsilon^2)$, $\epsilon_t= -\epsilon_{1,0}$,  $\epsilon =r/R$, prime denotes the radial derivative, $j_{0\parallel}$ is the equilibrium plasma current.

Let us multiply equation (\ref{eq1})  by $\Phi_{mn}^*$ and integrate the product over the plasma volume. Like in subsection \ref{subsec:1}, we apply a perturbative approach in order to obtain an equation for the  damping / growth rate. As a result, we will have:
 \begin{eqnarray}\fl
2 \gamma \mathcal{W}\equiv  \sum_{mn}\gamma \int d^3x \frac{c^2\delta_0}{4\pi v_A^2}\left \{\left (1 -\frac{d\omega_{G}^2}{d\omega^2} \right )|\Phi_{mn}^\prime|^2 +
\frac{m^2}{r^2}\left (1 -\frac{d\omega_{G}^2}{d\omega^2} \frac{r^2\epsilon^{\prime 2}_t}{\epsilon_t^2}  \right )|\Phi_{mn}^2| \right \}\nonumber \\
= 0.5\sum_{mn}\mbox{Re}\int d^3 x \left (\mathbf{j}_{\perp mn}^\alpha + \mathbf{j}_{\perp mn}^{kin}\right )\cdot \nabla_\perp \Phi^*_{mn}. 
  \label{W2}
 \end{eqnarray}
   Here $\mathcal{W}$ differs from $\mathcal{W}_A$ by the presence of the term produced by coupling of $\tilde{\Phi}$ and $\tilde{\zeta}$.  When deriving this equation it was taken into account that the product $\omega B_{mn}^r$ (with $B_{mn}^r$ expressed through $\Phi_{mn}$) does not depend on $\omega$. This equation agrees with equation ($\ref{gamma}$) but it takes into account the plasma compressibility. 

 The first term in the RHS of this equation describes the instability drive by the energetic ions, $\gamma_\alpha$. The second one describes the damping,
 $\gamma_d$ (if plasma is equilibrium). Relations for  $\gamma_\alpha$  and $\gamma_d$ can be written as follows:
\begin{equation}
\gamma_\alpha =\frac{1}{4\mathcal{W}}\sum_{mn}\mbox{Re} \int d^3x \left (\mathbf{j}_{\perp mn}^\alpha \right )\cdot \nabla_\perp \Phi^*_{mn}
  \label{ga}
 \end{equation}
and
\begin{equation}
\gamma_d = \frac{1}{4\mathcal{W}}\sum_{mn}\mbox{Re}  \int d^3x \left (\mathbf{j}_{\perp mn}^{kin} \right )\cdot \nabla_\perp \Phi^*_{mn},
  \label{gd}
 \end{equation}
where $\mathcal{W}$ is defined by equation (\ref{W2}).
The instability growth rate is $\gamma_\alpha +\gamma_d$, with  $\gamma_\alpha>0$ and $\gamma_d<0$. 

\subsection{Mode damping / growth  rate in  a Maxwellian plasma with a beam}
\label{subsec:3}

  Following the procedure described in \cite{isomon} but taking into account various harmonics  in the field line curvature and  assuming Maxwellian the velocity distribution of the bulk plasma  we obtain:
\begin{equation}\fl
j_{r(mn)}=\frac{iMc^2}{4\bar{B}^2r^2}\sum_{\mu\nu} (\mu^2\epsilon_{\mu\nu}^2\Phi_{m,n}^\prime
-\mu m\epsilon_{\mu\nu}^\prime \epsilon_{\mu\nu}\Phi_{m,n} )\int d^3v \frac{w^4}{\omega -k_{m+\mu,n+\nu N}v_\|}\hat{\Pi}F,
\label{jr2}
 \end{equation}
\begin{equation} \fl
j_{\vartheta(mn)}=\frac{Mc^2}{4\bar{B}^2r}\sum_{\mu\nu}\epsilon_{\mu\nu}^\prime (\mu\epsilon_{\mu\nu}\Phi_{m,n}^\prime
-m\epsilon_{\mu\nu}^\prime \Phi_{m,n} )\int d^3v \frac{w^4}{\omega -k_{m+\mu,n+\nu N}v_\|}\hat{\Pi}F,
\label{jt2}
 \end{equation}
 where   $w^2 =(0.5 v_\perp^2+v_\|^2)$, $\hat\Pi$ in the ($r, v$) variables ($r$ is approximately a constant of motion for well-passing particles) is
 \begin{equation}
\hat{\Pi}= -\frac{2}{v_{T}^2}+\left (\frac{\omega}{\omega_\varphi} +n\right )\frac{1}{\iota\omega\omega_B}
\frac{1}{r}\frac{\partial}{\partial r},
	\label{Pi0}
 \end{equation}
where $\omega_\varphi = v_\parallel /R$.  Note that these equations are valid for both ions and electrons. 

We assume that the second term in equation (\ref{Pi0}) is small and neglect it, which is justified when 
\begin{equation}
\frac{v_{T\sigma}}{2\omega r}|m+\mu -\nu N\iota^{-1}| \ll \frac{L}{\rho_\sigma},
	\label{neglect}
 \end{equation}
where $L= |d\ln F /dr|^{-1}$ is a characteristic length of the plasma inhomogeneity,  $\rho_\sigma =v_{T\sigma}/\omega_{B\sigma}$, $\sigma$ labels particle species (electrons and ions).
When obtaining equation (\ref{neglect}) we used the resonance condition (\ref{res1}).

 In the case, when  the second term in equation (\ref{Pi0}) exceeds the first one, the instabilities driven by the spatial inhomogeneity of the bulk plasma can arise in stellarators,
i.e., thermal particles can lead to instabilities in the same way as energetic ions do it. This was shown  for a TAE instability in W7-X in reference \cite{Koenies}. One can see, for instance, that   the helicity-induced resonance ($\mu=1,\,\nu=1$) leads to  $(L/\rho_i)_{cr} =21$   for a TAE mode with $m=5$ localized 
around the radius $r/a \sim 0.5$ in a plasma with $\beta_i =0.01$ and $\iota =0.9$.  Therefore, this mode can be destabilized provided that $L/\rho_i < 21$.  Of course, whether the mode will be destabilized or not depends on the power balance in the region where 
$L/\rho_i<(L/\rho_i)_{cr}$ and the region where $L/\rho_i>(L/\rho_i)_{cr}$  within the mode width. 
Note that $L$ is strongly enlarged in Maxwellian plasmas due to the temperature inhomogeneity when $v_\|^{res}>v_T$,
where $v_\|^{res}$ is defined by equation (\ref{res1})~\cite{Alfvegrad}.

We are interested only in the imaginary parts of the integral in these equations, which arise due to resonance ($\ref{res1}$). 
Replacing  $1/\Omega $ with $-i\pi \delta (\Omega)$ we obtain for a plasma with Maxwellian distribution  (c.f. references \cite{VanDam,BKS,AI} where Maxwellian distribution for hot ions was used to study the destabilization of TAE modes in tokamaks):
 \begin{equation}
\mbox{Im}\int d^3v \frac{w^4}{\omega -k_{m+\mu,n+\nu N}v_\|}\hat{\Pi}F= \frac{\sqrt{\pi} n_\sigma\omega}{k^2_{res}}Q(u), 
	\label{Pi3}
 \end{equation}
where   
 \begin{equation}
Q(u)=\frac{1}{u}(2u^4+2u^2+1)e^{-u^2}-Q_\epsilon ,
\label{Q1}
\end{equation}
\begin{equation}
Q_\epsilon = \frac{1}{u}\left [2\left (1+\frac{1}{2\epsilon_{ef}}\right )^2 u^4 +  2\left (1+\frac{1}{2\epsilon_{ef}}\right )u^2  +1\right ]
e^{-\left (1+\epsilon_{ef}^{-1}\right )u^2},
\label{Q2}
\end{equation}
$k_{res} \equiv k_{m+\mu , n+\nu N}$,  $n_\sigma$ is the particle (ion or electron) density, $u=|v^{res}_\||/v_{T}$, $\epsilon_{ef}=\sum_{\mu\nu} |\epsilon_{\mu\nu}|$ ($\mu =\nu \neq 0$) is an effective Fourier harmonic of the magnetic field which determines the boundary for the  well passing particles in the velocity space region. In particular, in the W7-X high-mirror configuration, $\epsilon_{ef}$ varies from $0.08$ at the magnetic axis to $0.24$  near the plasma edge.  In calculation, we assumed that passing particles have transverse velocities determined by $0<v_\perp < |v_\parallel |/\sqrt{\epsilon_{ef}}$. The term $Q_\epsilon$ decreases the damping, but it is considerable only when 
$u \lesssim \sqrt{\epsilon_{ef}} $. The function $Q(u)$ is rather flat in the region $0.5 \lesssim u\lesssim 1$, which facilitates making  estimates in the case when   the ratio $\omega /(|k_{res}|v_T)$ lies in this interval but the mode frequency is not known. On the other hand, when the frequency is known one can use the function $u^2Q(u)$ which arises  when $k_{res}^2$ in the RHS of (\ref{Q2}) is eliminated by means of relation  $k_{res} =\omega /v_\|^{res}$.   
	
\begin{figure}
\centering
\includegraphics[width=0.8\columnwidth]{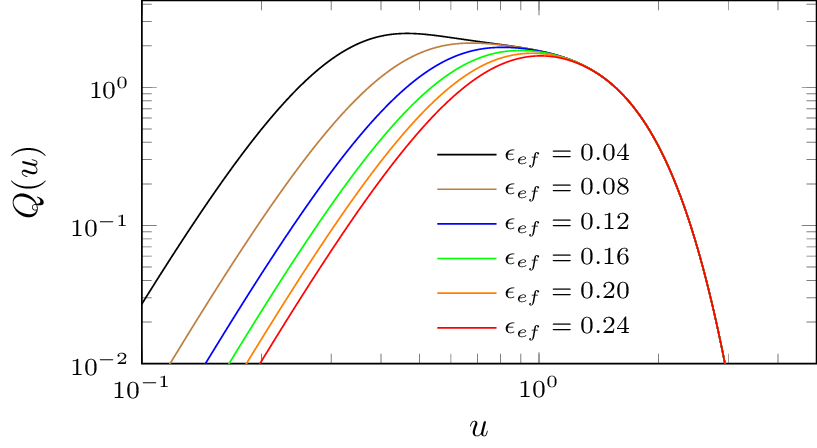} 
\caption{$Q$ versus $u$ for various $\epsilon_{ef}$, with $u=|v_\|^{res}|/v_T$.  In the W7-X  high mirror configuration, $\epsilon_{ef}$ varies from $0.08$ at the magnetic axis to $0.24$ near the plasma edge; in the standard configuration $0.03\lesssim \epsilon_{ef}\lesssim 0.18$.  }
\label{Fig:Q}
\end{figure}

Knowing components of the current $\mathbf{j}^{kin}_{mn}$  and using equations  (\ref{gd}), (\ref{W2}) with $\mu = 0,\pm 1$, we can write the following equation for the damping rate: 
 \begin{equation}\fl
 \frac{\gamma_d^{(\sigma )}}{\omega} = -\frac{\sqrt{\pi}}{8\delta_0}\frac{M_\sigma}{M_i}\frac{\sum_{mn}\int_0^a dr r n_\sigma (r) \sum_{\mu\nu}\epsilon^{-2}
\left |\mu\epsilon_{\mu\nu} \Phi_{mn}^\prime - m\epsilon_{\mu\nu}^\prime \Phi_{mn}\right |^2Q (u_\sigma)\bar{k}^{-2}_{res}}
{\sum_{mn}\int_0^a dr r^{-1}n_i(r)  \left (g_1 r^2|\Phi_{mn}^\prime|^2 +g_2m^2|\Phi_{mn}|^2\right )},
	\label{gd3}
 \end{equation}
where  $\epsilon =r/R$, $n_\sigma$ is the particle density,  $\sigma =e,\,i$ labels  electrons and ions,  
 $\bar{k}_{res} \equiv k_{res}R=(m+\mu)\iota -(n+\nu N)$,
$u=R\omega /(|\bar{k}_{res}|v_{T})$,
$g_1= 1-d\omega_G^2/d\omega^2$, 
$g_2= 1-(d\omega_G^2/d\omega^2)(r^2\epsilon_t^{\prime 2}/\epsilon_t^2 )$. The function $Q(u_\sigma)$ is shown in figure \ref{Fig:Q}.  Fourier harmonic $\epsilon_t$ is approximately proportional to $r$; therefore, $g_1\approx g_2$ and  $g_1$ can be written as
	 \begin{equation}
g_1 = 1 +\tilde{\epsilon}^2\sum_{l=\pm 1} \frac{k_{m+l,n}^2 c_s^4}{R^2(\omega^2-k^2_{m+l,n}c_s^2)^2}. 
\label{g11}
\end{equation}
The same relation  for $g_1$   can be obtained from equation (\ref{eq1})  by keeping only the term proportional to  $\Phi_{mn}^{\prime\prime}$ after calculation of the imaginary part of $j_{r,mn}^{kin}$ (which contributes to this term).
We assume that $\omega^2 \neq k^2_{m+l,n}c_s^2$, i.e., we do not consider  Alfv\'en-sound resonances and concomitant gaps in  Alfv\'en continuum
(the case of Alfv\'en-sound resonances deserves a special study, our analysis is not valid for it). 

Because
 $c_s^2 /v_A^2 \ll 1$ and   $\tilde{\epsilon}^2 \equiv \epsilon_t^2/(\delta_0\epsilon) \ll 1$, it is sufficient to use an approximate equation for $g_1$. For instance, approximating the mode frequency by  equation (\ref{gapfreq}), we can write equation  (\ref{g11}) in the case of TAE modes as follows:
\begin{equation}
g_1 = 1+4\tilde{\epsilon}^2\frac{c_{s*}^4}{\iota^2v_{A*}^4}\frac{R^2_\omega}{R^2}\sum_{l=\pm 1}\frac{(1+2l)^2}{1-(1+2l)^2c^2_{s*}/v_{A*}^2}, 
\label{gTAE}
\end{equation}
where the subscript ``*" labels magnitudes at the radius where the rotational transform is defined by equation (\ref{gapiota}). One can see that typically this $g_1$ is close to unity.

We observe in equation (\ref{gd3})  that the damping rate is proportional to $M_\sigma/M_i$, which may produce an illusion that the electron damping is always small. Therefore, we have to make some comments on his issue. 
  
	Let us assume first that  $|v_\parallel^{res}| =v_T^{\sigma}$. Then we obtain  from the resonance condition that 
	$\bar{k}_{res}^2 =\omega^2R^2/v_{T\sigma}^2 \propto M_\sigma$ for a given  mode frequency and, hence,  the damping does not depend on $M_\sigma$. This is a consequence of the fact that the drift velocity $v_D$ is determined by the particle energy, but not velocity.  In reality, however, the fixed magnitudes are the mode frequency and mode numbers, rather then $v_\parallel^{res}$. Therefore, assuming that $\omega$ and $\bar{k}_{res}$ are given,
we eliminate $M_\sigma$  in the expression for $\gamma_d^{(\sigma)}$ by means of the relation 
$\bar{k}_{res}^{-2}M_\sigma /M_i = u_\sigma^2c^2_\sigma /(R^2\omega^2)$, with $c_\sigma^2 =2T_\sigma /M_i$.  Now it is clear that the electron damping rate depends on the ratio $v_\|^{res}/v_T$, like the ion damping rate, there is no additional mass dependence.

Note that an expression similar to (\ref{gd3}) can be obtained for the instability drive by integrating (\ref{Pi3}) with the distribution function of fast particles, 
$F_\alpha$. 
For the beam ions, $F_\alpha$ can be approximated as \cite{isomon} 
\begin{equation}
F_\alpha =\frac{2n_b(r)}{\pi (1+\chi^2_0)v^3}\delta (\chi -\chi_0)\eta (v_0 -v), 
	\label{Fb}
 \end{equation}
where subscript "b" labels  beam particles, $\eta (v_0 -v)$ is the Heaviside step function, $\chi =v_\| /v$, the particle density is defined by $n_b =p_b /\mathcal{E}_0$,  $\mathcal{E}_0=0.5M_b v_0^2$,  $p_b$ is the energetic ion pressure defined by $p_b=0.5(p_\parallel + p_\perp)$,
$p_\parallel =\int d^3v v_\parallel^2 F_\alpha$, $p_\perp =0.5\int d^3 v v_\perp^2 F_\alpha$, and we take $M_b =M_i$. Equation (\ref{Fb}) implies that the ion  energy is sufficiently high, $\mathcal{E}\gg (M_i/M_e)^{1/3}T_e$, so that Coulomb collisions mainly slow down the fast ions without much pitch-angle scattering.   
 In this case, an equation for $\gamma_\alpha$  has the form of (\ref{gd3}) where the subsript $\sigma$ should be replaced by $b$,  $c_b^2 =2\mathcal{E}/M_b$,  and  $Q(u)$ replaced by  
\begin{equation}
Q_b (u_b)=-\frac{\sqrt{\pi}}{(1+\chi_0^2)}\left [\frac{\omega_{*b}}{\omega}\left (\frac{1}{\chi_0^2} +1\right )^2u_b^2
  +\frac{3}{\chi_0^4} - \frac{2}{\chi_0^2} -5 \right ],
	\label{Qb}
 \end{equation}
with $\omega_{*b} = n[1+\sigma_v\omega R /(nv_0u_b)]v_0^2(\iota\omega_{Bb}r)^{-1}\partial{\ln n_b}/{\partial r}$,  
$u_b =|v_\|^{res}| /v_0$, $\sigma_v =\mbox{sgn}\, v_\|^{res}$.

\subsection{Local approximation for damping rate and growth rate caused by the beam }
	\label{subsec:local}
	
	In order to calculate the damping rate  by means of  (\ref{gd3})  one has to know the radial profile  of the mode amplitude, $\Phi (r)$. However, simple estimates for $\gamma_d^{(\sigma)}$ can be made by using a local 
	approach. To construct local $\gamma_d^{(\sigma)}$ one has to   assume, for instance,  that terms in  equation (\ref{gd3}) containing the radial derivatives of $\Phi_{mn}$  dominates and the mode width is very narrow. Then  equation (\ref{gd3}) reduces to 
	 \begin{equation}
 \frac{\gamma_d^{(\sigma ) \,loc}}{\omega} = -\frac{\sqrt{\pi} }{8\delta_0 g_1^2}\frac{M_\sigma n_\sigma}{M_in_i}\sum_{\mu\nu} 
\left (\frac{\mu\epsilon_{\mu\nu}}{\epsilon}  \right )^2 \bar{k}_{res}^{-2}
Q(u_\sigma),
	\label{loc1}
 \end{equation}
where all the magnitudes are taken at the radius where the mode amplitude is maximum.

Following \cite{isomon}, let us introduce  the mode dimensionless frequency, $\bar{\omega} =\omega R /c_{s0}$, with $c_{s0} =c_s(0)$. Then 
$u =\bar{\omega}c_{s0}/(|\bar{k}_{res}v_T|)$, which reduces to   $u = \bar{\omega}\sqrt{\Gamma}/(|\bar{k}_{res}|\sqrt{\Theta})$ with
 $\Theta =T(r)/T_0$ for a plasma with $T_i=T_e$, and  
 \begin{equation}
g_1 = 1 +\tilde{\epsilon}^2\sum_{l=\pm 1} \frac{\Theta^2\bar{k}_{m+l,n}^2 }{\bar{\omega}^4(1-\bar{k}^2_{m+l,n}\Theta/\bar{\omega}^2)^2}. 
	\label{g111}
 \end{equation}
 It follows from (\ref{g11}) and ($\ref{g111})$ that the  plasma compressibility increases the mode energy  ($g_1>0$) and, thus, decreases $\gamma_d$. 
However, the effect is small, $g_1\approx 1$, when  $\omega^2 \gg k_{m+l,n}^2c_s^2$.  

The damping rate (\ref{loc1}) depends on the mode numbers.  It  may be preferable to have an expression for $\gamma_d$ containing the mode frequency instead of mode numbers.   Eliminating $\bar{k}_{res}$, we obtain: 
  \begin{equation}
  \frac{\gamma_d^{(\sigma ) \,loc}}{\omega} = -\frac{\sqrt{\pi} }{8g_1^2\delta_0}\frac{n_\sigma}{n_i}\sum_{\mu\nu} 
	\left (\frac{\mu\epsilon_{\mu\nu}}{\epsilon}  \right )^2 
 \frac{c_\sigma^2}{\bar{\omega}^2 c^2_{s0}}  u_\sigma^2Q(u_\sigma).
	\label{loc2}
 \end{equation}
This equation and (\ref{loc1}) are valid for both electrons and ions. When $T_e=T_i$, the electron damping equals the ion damping  for 
$u_e^2Q(u_e) =u_i^2Q(u_i)$.  In particular,  $\gamma_{d,e}^{loc} =\gamma_{d,i}^{loc}$  for  $u_e \approx 0.1$ and $u_i\approx 4$ in a hydrogen plasma and 
 $u_e \approx 0.07$ and $u_i\approx 4.2$ in a deuterium  plasma.

A similar equation is valid for the fast ions. Due to this we can write a simple estimate for the threshold density of fast ions (for which the system is on the margin of stability). Assuming that there is only one dominant harmonic of the magnetic field in the damping  rate and another one in the growth rate ($\gamma_\alpha$),  we obtain: 
 \begin{equation}
 \frac{n_b^{cr}}{n_\sigma} = \frac{T_\sigma}{\mathcal{E}_0 }\left (\frac{\mu_\sigma\epsilon_{\mu\nu}^{(\sigma)}}
{\mu_b\epsilon_{\mu\nu}^{(b)} }\right )^2\frac{\sum_j u_\sigma^2 Q(u_\sigma)}{\sum_j u_b^2Q_b(u_b)}.
	\label{crit}
 \end{equation}
Here  $\sigma =e, \,i$; $j=\pm\mu$, $\mbox{sign}\,\mu =\mbox{sign}\,\nu$.

Equations obtained above in the local approximation  do not include a contribution of the mirror harmonic, $\epsilon_{01}$, because the  term 
containing $\Phi_{mn}^\prime$ in (\ref{gd3}) is proportional  $\mu$.   The effect of the mirror harmonic  can be evaluated   by taking   
$\Phi_{mn} (r) \propto \exp (ik_r r)$ and neglecting the radial dependence in other magnitudes, which leads to 
	 \begin{equation}
 \left .\frac{\gamma_d^{(\sigma ) \,loc}}{\omega}\right |_{\mu =0} = -\frac{\sqrt{\pi} }{8\delta_0g_1^2}\frac{M_\sigma n_\sigma}{M_in_i}
\frac{k_\theta^2}{k_\perp^2}\left(\frac{r\epsilon_{01}^\prime}{\epsilon} \right )^2 \bar{k}_{res}^{-2}Q(u_\sigma),
	\label{loc0}
 \end{equation}
where $k_\perp^2 =k_r^2 + k_\vartheta^2$. For $\mu \neq 0$, this procedure leads to equation (\ref{loc1}) due to the relation  
$r\epsilon_{\mu\nu}^\prime =\epsilon_{\mu\nu} $. Using the relation $\bar{k}_{res}^2 \bar{\omega}^2=u^2_\sigma v_{T\sigma}^2/c_{s0}^2$ we obtain an expression similar to (\ref{loc2}), but for mirror harmonic. Comparing it with (\ref{loc2}) we conclude that due to the mirror harmonic the damping /growth rate increases by the factor  
\begin{equation}
\mathcal{F}=1+\frac{k_\vartheta^2}{k_\perp^2}\left (\frac{r\epsilon_{01}^\prime }{\mu\epsilon_{\mu\nu}}\right )^2. 
\end{equation}
To  evaluate this magnitude, we take $\mu=\nu =1$ and 
 $\epsilon_{01} = \epsilon_{01}(0)(1+\zeta r^2/a^2)$, where $\epsilon_{01}(0) =0.09$ and $\zeta=0.33$ for the considered W7-X high mirror configuration with $\beta (0) =0.037$. Then $\mathcal{F} = 1+0.55 (k_\vartheta r)^2/ (k_\perp a)^2 < 1.55$. Therefore, the mirror harmonic weakly contributes to the damping /growth rate.

\section{Specific examples}
\label{sec:examples}

\subsection{Damping and growth rates of isomon modes in Wendelstein 7-X}
\label{subsec:iso}

This subsection is devoted to the study of damping rates of isomon modes in the first planning NBI experiments on W7-X.  General relations derived above and  the eigenmodes found in the work \cite{isomon} will be used in our calculations. 
Note that because the mirror harmonic weakly contributes to $\gamma$,  whereas other harmonics are approximately the same in the high mirror configuration and standard configuration, the damping rate and growth rate  are roughly equal in these configurations.

 Let us begin with the local approach. Assuming that the temperature profile is given by figure 3 of \cite{isomon} we take 
 $\Theta \equiv T(r)/T_0= 0.77$, which corresponds to $r/a\sim 0.7$ where the mode amplitudes are close to their maxima. Then, using equation (\ref{vr}), we obtain for the $\mu /\nu =1 / 1$ helical resonance:  
 \begin{equation}
u_i  \equiv |v_\|^{res}/v_{Ti}| =\frac{1.47\bar{\omega}}{|m\Delta \iota \pm (5-\iota)|}.
	\label{uloc}
 \end{equation}
For instance, for  $m=3$, $\iota =0.9$,  and $\bar{\omega} = 1.98$ this yields $u_i=0.65$ and $0.76$.  For $m=4$, $\iota =0.9$, and $\bar{\omega} = 2.56$ this yields $u_i=0.83$ and $1.02$. Then $Q(u_i)\approx 2$, as follows from figure \ref{Fig:Q}.
In addition, we  obtain $\bar{k}_{res} = 4.4$ and $3.8$  for $m=3$;   $\bar{k}_{res} = 4.5$ and $3.7$  for $m=4$.
 Using these magnitudes and 
taking $\epsilon_{11}^2/\epsilon^2 =0.8^2$,  $\delta_0 =1.5$,  we obtain from equation (\ref{loc1}) that the ion damping is  $|\gamma_{d,i}^{loc}| /\omega\approx 0.02$ for both
 the $m=3$  and $m=4$ modes.

\begin{figure}
\centering
\includegraphics[width=0.7\columnwidth]{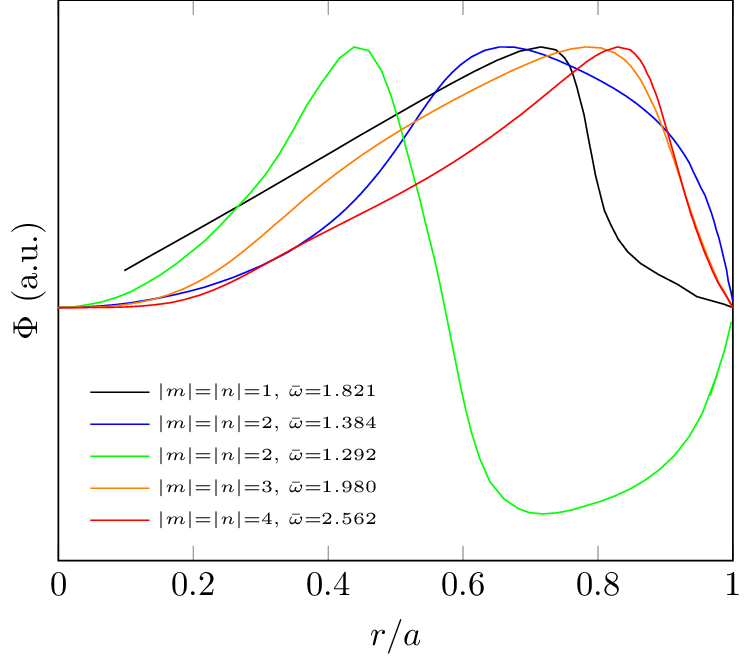} 
\caption{Isomon modes in Wendelstein 7-X \cite{isomon}.}
\label{Fig:modes}
\end{figure}
\begin{table}[h]
\centering
\caption{Damping rates calculated by means of equation (\ref{gd3}) with $\delta_0=1.5$ for the IM modes shown in figure \ref{Fig:modes}. }
\label{Table:Tie0}
\resizebox{0.3\textwidth}{!}{
\begin{tabular}{|l|c|c|}
\hline
m=n  & $\bar{\omega}$& $|\gamma_d|/\omega$  \\ \hline
1 & 1.821 & 0.0133 \\ \hline
2 & 1.384 & 0.0239 \\ \hline
2 & 1.292 & 0.0263 \\ \hline
3 & 1.980 & 0.0199 \\ \hline
4 & 2.562 & 0.0192 \\ \hline
\end{tabular}}
\end{table}

On the other hand, more realistic calculations based on equation (\ref{gd3}) and figure \ref{Fig:modes} for the modes with $m= 1 -4$ are shown in Table \ref{Table:Tie0}.
We observe that the damping rates   in  this Table and  the results of the local approach are in good agreeement.
 In addition, we observe that the damping rates of all the considered modes are roughly equal.

The electron damping  is small or, at least, it cannot be described by our relations  because $u_e \ll 1$ and, thus, trapped particles can be responsible for the electron damping. 
	
Now we proceed to consideration of the instability drive, $\gamma_\alpha$, by using equation (\ref{gd3}) with $Q_b$ given by (\ref{Qb}). Energetic ions at the initial stage of operation  with NBI are described in Appendix A  of work \cite{isomon}. As shown in figure A2 of the mentioned work,  at each radius there are two sharp maxima in 
the $\lambda$ distribution of injected particles ($\lambda =\mu_p \bar{B}/\mathcal{E}$ is the particle pitch parameter, $\mu_p$ is the particle magnetic moment), but only one of them  is relevant to passing particles; almost no passing ions were produced  at   $r/a > 0.8$. Moreover,  well passing particles were produced only in the plasma core, mainly at $r/a < 0.4$ where $\lambda \approx 0.75$ ($\chi_0 \approx 0.5$). This means that only a core region contributes to the numerator  of  (\ref{gd3}).  This considerably reduces the growth rate of isomon modes because their  maximum amplitudes lies  at $r/a > 0.5$.   

\begin{figure}
\centering
\includegraphics[width=0.6\columnwidth]{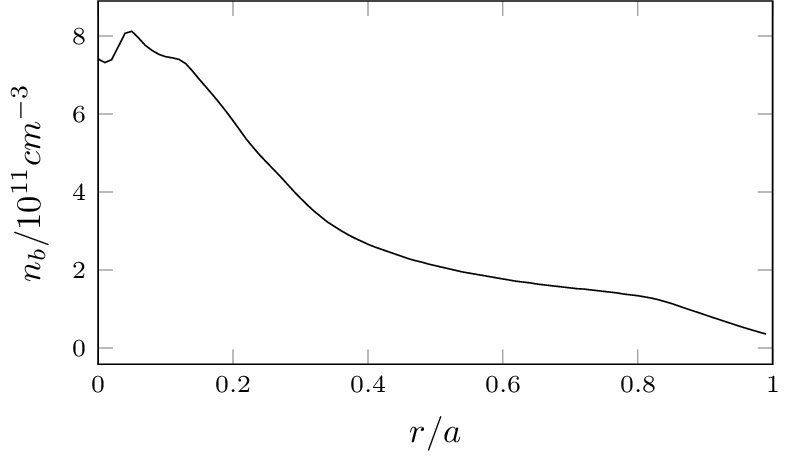} 
\caption{Radial distribution of NBI ions, which was used in calculation of the growth rate of the IM instability.  }
\label{Fig:profile}
\end{figure}

\begin{figure}
\centering
\includegraphics[width=0.7\columnwidth]{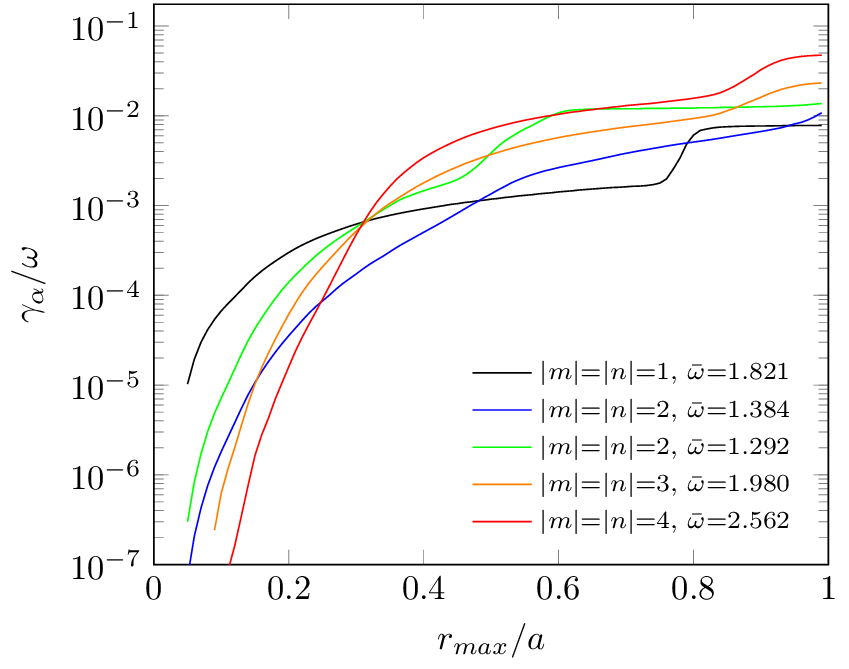} 
\caption{Dependence of the IM instability drive produced by NBI ions in W7-X on the maximum radius, $r_{max}$, restricting the region where well passing NBI ions are located. Calculations were carried out for protons with $\mathcal{E} =55$ keV and 
$|\chi_0 | = 0.5$. Because most well passing injected ions are located in the region $r/a<0.4$\cite{isomon}, we conclude
 that $\gamma_\alpha /\omega \sim 10^{-3}$. }
\label{Fig:drive}
\end{figure}

Taking this into account and assuming that the radial distribution of the energetic ions coincides with the energy deposition profile  of these ions (shown in figure \ref{Fig:profile}),  we calculated $\gamma_\alpha$. We found that $\gamma_\alpha /\omega \sim 10^{-3}$ when the upper limit in the numerator of (\ref{gd3}) is $0.4 a$, see figure \ref{Fig:drive}. This implies that the drive  produced by well passing NBI ions is not sufficient to overcome the damping. The role of the NBI ions in the region $r/a >0.4$ is not clear because they are mainly marginally passing  and transitioning, which are not described by our theory. 
 
Note that although the mode amplitude is maximum in the region where the number of well passing NBI ions is very small, 
  it is possible to evaluate $\gamma_\alpha$ by modifying local equation (\ref{loc1})  as follows:
 \begin{equation}
 \frac{\gamma_\alpha^{(loc)}}{\omega} = -\frac{\sqrt{\pi} }{8\delta_0 g_1^2}
\frac{M_b n_b}{M_in_i}\left (\frac{\epsilon_t}{\epsilon}  \right )^2\mathcal{K}_{1}\mathcal{K}_{2}
\sum_{\mu =\pm 1}  \bar{k}_{res}^{-2}Q_b(u_b),
	\label{locdrive}
 \end{equation}
where $\mathcal{K}_{1}\approx 0.5$ is the fraction of NBI ions from those injectors which produce mainly passing ions,  
 $\mathcal{K}_{2} =|\Phi_{0.3}|^2/{|\Phi_{max}|^2}$,  $\Phi_{0.3}$  and $\Phi_{max}$ is the mode amplitude at $r/a =0.3$ and at the radius where it is maximum, respectively; $\mathcal{K}_{2}$ can be obtained from  figures 4  and 6 of of reference\cite{isomon}: $\mathcal{K}_{2} =0.02$ for the mode with $m=n=2$,  
$\mathcal{K}_{2} =0.04$ for the modes with $m=n=3$ and $m=n=4$.  Equation (\ref{locdrive}) yields growth rates which are in qualitative agreement with those shown in figure \ref{Fig:drive} for $r_{max}/a =0.4$.

\subsection{Damping of TAE modes in Wendelstein 7-X}
\label{subsec:W7X}
According to section \ref{subsec:rgm}, the ratio of the $\mu /\nu =1 /1$  resonance velocity to the ion thermal velocity  during TAE instabilities   in the first NBI experiments  on W7-X can be about unity. Therefore, one can expect that TAE damping  will be strong, like in the case of isomon modes.

 Let us first  make a simple estimate by using equation (\ref{loc1}) for a plasma with the same parameters as in the previous section. For $\iota_*=0.9$ we obtain  $\bar{k}_{res} =4.55$ and $3.65$, which leads to 
$u=1$ and  $1.23$, see figure \ref{Fig:Q}. Using equation (\ref{loc1})  we obtain then that $|\gamma_d^{loc}| /\omega \sim  0.02$.

\begin{figure}
\centering
\includegraphics{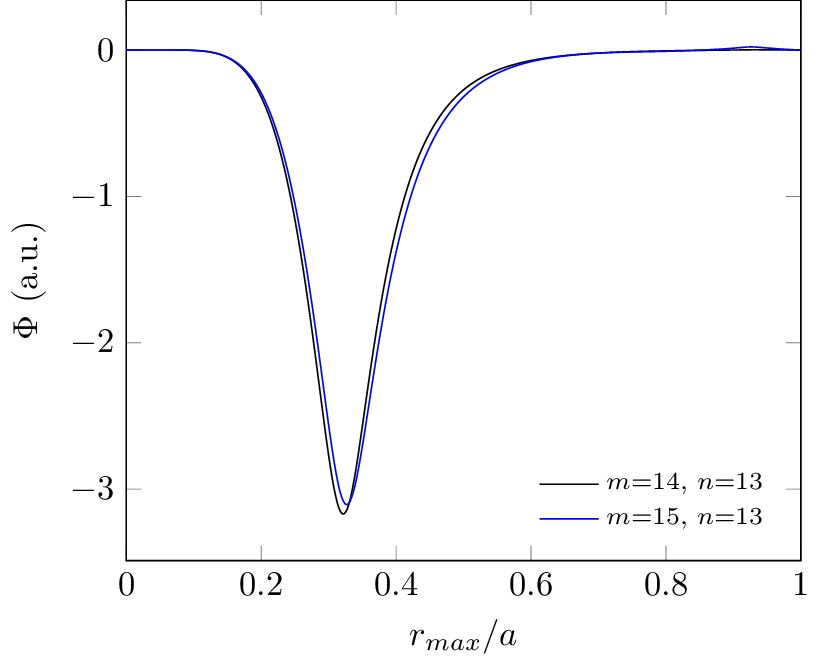} 
\caption{A TAE mode in Wendelstein 7-X.
 The mode frequency equals $42.48$ kHz, dominant  mode numbers are $m=14,\,\,15$ and $n=13$~\cite{KoeniesPC}.}
\label{Fig:TAE}
\end{figure}

 As an example, we consider the TAE mode with $m=14,\,\,15$ and $n=13$, see figure \ref{Fig:TAE}, with the mode frequency $42.48$ kHz~\cite{KoeniesPC}.
Using equation  (\ref{gd3}) we find that the damping rate is $|\gamma_d |/\omega=0.0244$.
The damping rate weakly depends on the particle density  and temperature profiles because the mode width is rather small: it does not change appreciably
if we take $n_e=\mathrm{const}$ or $T_i=T_i(r_*)=\mathrm{const}$.
Artificially setting the mode frequency to values as high as $70$ kHz gives the damping rates shown
 in Table (\ref{Table:Tae1}). All these magnitudes are in qualitaive agreement with the local estimate above. 
\begin{table}[h]
\centering
\caption{Damping rates of a TAE mode with the realistic frequency ($42.48$ kHz) and artificially increased frequencies in W7-X. Calculations
 were  
made  by using equation (\ref{gd3}) with $\delta_0=1.5$ for the mode structure shown in figure \ref{Fig:TAE}. }
\label{Table:Tae1}
\resizebox{0.3\textwidth}{!}{
\begin{tabular}{|l|c|c|}
\hline
$f$,kHz& $|\gamma_d|/\omega$  \\ \hline
70     &    0.0152 \\ \hline
60     &    0.0192 \\ \hline
50     &    0.0225 \\ \hline
42.48  &    0.0244 \\ \hline
\end{tabular}}
\end{table}

\subsection{Damping of TAE modes in an  LHD experiment}
\label{subsec:LHD}
Our analysis above predicts a rather large  Landau damping of isomon modes and TAE modes in the Wendelstein 7-X.   On the other hand,  TAE instabilities were already observed in helical systems, in particular, in the LHD device, in the discharge 
\#24512 \cite{Toi,Yamamoto}. Below we evaluate  Landau damping for this LHD discharge  in order to see whether our theory is consistent with the experiment.

In the mentioned discharge an Alfv\'enic activity  with $n=1$  and $n=2$ was observed with the frequencies in the range of 50-80 kHz during tangential NBI with the particle (protons) energy $\mathcal{E}_0 $ up to $150$ keV injected into a helium plasma 
\cite{Yamamoto}. The eigenmode calculations carried out with the BOA code \cite{AE} have found two discrete  TAE modes with $n=1$ \cite{LHD}. One of them  was "even" TAE mode with the $50$ kHz frequency, another one with $60$ kHz was "odd" TAE. Both modes were localized around the $r/a \sim 1/3$ flux surface,
  $\gamma_\alpha /\omega  \approx 0.3$ for the even mode and   $\gamma_\alpha /\omega  \approx 0.03$ for the odd mode were calculated \cite{LHD}.  
	
	In the core region, the beam beta well exceeded the plasma beta: $\beta_b (0) =1.7 \% $ , $\beta (0) \approx 0.45\%$.
	The electron density at the magnetic axis was $n_e (0) = 10^{19}$ m$^{-3}$. 
Thus, beam-plasma parameters in LHD were very different from those which are expected in the first NBI experiments on Wendelstein 7-X.  
Key differences are a bigger fast-ion population and a lower plasma beta. In addition, in LHD  the number of the field periods is  larger by a factor of two 
($N_{LHD}=10$) but the  rotational transform in the core region smaller  ($\iota_{LHD}=0.4 - 0.5$). Because of this, the largest $\mu/\nu = 2/1$ gap in the AC was very far from the TAE gap  (the ratio $\omega_{*,21}/\omega_{*,10} =23$, which is much larger than in W7-X). 
 Due to this, the effect of this helical gap on the shift of the TAE gap was small in the core region ($r/a <0.4$),
see figure 7a in \cite{LHD}.  In spite of these differences, $\beta_i$ was about that given by equation (\ref{beta}),  like in W7-X. 
For this reason, Landau damping was not small.

We restrict ourselves with a local relation to calculate the damping rate. As shown in previous subsections, this approximation gives the results which are in qualitative agreement with those obtained by using  a more rigorous equation for $\gamma_d$.    
Because in the region of location of the modes ($r/a \sim 0.3$) the observed frequency $\hat{\omega}$ lay inside the TAE gap 
 (see figure 7a in \cite{LHD}),  we can take $R_\omega =R$.  In addition,
  the helical harmonics  $\epsilon_{11}$ and $\epsilon_{21}$  in LHD were approximately equal (see figure 6 in \cite{LHD}). Therefore,  we have to take into account both of them. They lead   to $v_\|^{res} /v_{A*}=(1\pm 36)^{-1}$ and  $v_\|^{res} /v_{A*}=(1\pm 38)^{-1}$, respectively, at $\iota =0.5$, and to smaller magnitudes at $\iota =0.4$ (because the magnetic shear in LHD is not small, the iota varies considerably within the mode location).  This leads to $u = 0.55 - 0.61$ for $\beta_i =\beta_e = 2.14 \times 10^{-3}$. At 
 $r/a =0.3$ we take $\Theta \equiv 
T_i(r)/T_0 =0.9$ and $\epsilon_{ef} =0.08$, for which $Q(u) \approx 2$. The normalized frequency is $\bar{\omega} =2.7$ for the even mode and  $3.24$ for the odd mode.  The ratio $\epsilon_{\mu\nu} /\epsilon$  for $\epsilon_{11}$ and $\epsilon_{21}$ can be evaluated as 
$0.5$. Then equation (\ref{loc1}) yields $|\gamma_d^{loc} |/\omega = 0.01 $ for the even mode and $7.6 \times 10^{-3}$ for the odd mode, which is much less than $\gamma_\alpha /\omega$ due to a very large pressure gradient of the beam particles in the region where TAEs are located.

The estimates made are rather rough because the fraction hydrogen  in the helium plasma was considerable but not known exactly. Nevertheless, they are sufficient to  conclude that in the considered LHD experiment the damping was weaker but the drive was much stronger than those expected in the first W7-X NBI experiments. 
   
	\subsection{Damping of high frequency  modes in Wendelstein 7-X}
As shown in section \ref{subsec:rgm}, $v_{i\|}^{res}/v_{Ti} \gg 1$ for HAEs and MAE modes, which implies that the ion Landau damping of these modes is exponentially small (unless $\beta_i$ is large). Therefore, below we consider the electron damping at small beta  and ion damping at large beta.

In order to evaluate electron  damping rates of high frequency gap modes we approximate their frequencies by  equation (\ref{gapfreq}). Then equation (\ref{loc2}) will take the form:
  \begin{equation}
  \frac{\gamma_d^{(e)\,loc}}{\omega} = -\frac{\sqrt{\pi} \beta_{e*}}{2g_1^2\delta_0}
	\frac{1}{(\mu_0\iota_*-\nu_0 N)^2}\sum_{\mu\nu} \left (\frac{\mu\epsilon_{\mu\nu}}{\epsilon}  \right )^2 u_e^2Q(u_e),
	\label{loc3}
 \end{equation}
where 
\begin{equation}
 u_e=\left (\frac{n_eM_e}{n_iM_i\beta_{e*}}\right )^{1/2} \left |1+2\frac{\mu\iota_* -\nu N}
{\mu\iota_0 -\nu_0 N }\right |.
	\label{ue}
 \end{equation}

 Let us see the influence on the damping of the toroidal harmonic of the magnetic field ($\mu =\pm 1$,  $\nu =0$). The analysis will be carried out for a hydrogen plasma with $\beta_{e*} =0.01$ and $\delta_0 =1.5$ in W7-X. Assuming $R_\omega =R$,   we obtain:  $u_e = 0.41$ and $0.16$ for HAE$_{11}$; $u_e = 0.525$ and $0.15$ 
for HAE$_{21}$. Because these  magnitudes of $u_e$, $Q(u_e)$  strongly depends on $\epsilon_{ef}$, as follows from figure \ref{Fig:Q}. This means that the damping is very sensitive the mode spatial location and the location of the mode frequency in the AC gap. In the high mirror case the damping is minimum, 
$|\gamma_d^{(e)loc}|/\omega \sim 5\times 10^{-6}$ for HAE$_{11}$ and $2\times 10^{-5}$ for HAE$_{21}$. For the modes with higher frequencies ($R_\omega <R$), the damping  rate is larger.  
 
We conclude from here that the effect  Landau damping associated with the $\mu/\nu =1/0$  resonance on high frequency gap modes is  smaller than that for TAEs and isomon modes. This is not surprising: in Wendelstein 7-X the toroidal harmonic is less by a factor of two than the helical harmonic (which yields the difference by a factor of four); in addition,  $u_e^{TAE} \ll u_i^{HAE,MAE}\lesssim 1$, which leads to a decrease of  $Q$.

Let us consider now the ion damping of MAE modes due to the helical resonance ($\mu=\nu = 1$). 
One can see that $\bar{k}_{res}= 6.6$ and $1.6$. Assuming that the mode is core localized where the pressure is high, 
$\beta_i=4\%$, we obtain $u_i = 1.9$ and $7.8$ for the larger and smaller  $\bar{k}_{res}$, respectively. Then equation (\ref{loc1}) 
yields $|\gamma_d| /\omega =10^{-3}$.   

	\subsection{Damping of Alfv\'en gap    modes in a Helias reactor}
	
	As shown in subsection \ref{subsec:rgm}, resonances with $\mu=\mu_0$, $\nu=\nu_0$ lead to  $v_\|^{res} =v_{A*}/3$ that provides  interaction of gap modes with thermal ions in high beta plasmas. In the Helias reactor dominant Fourier harmonics of the magnetic field  are the same as in the Wendelstein 7-X  high mirror configuration. Therefore, the resonance with  $v_\|^{res} =v_{A*}/3$ can  play an important role in damping of TAE modes and HAE$_{11}$ modes. One can see that 
	$|\bar{k}_{res}| =1.5 |\mu -\nu N |$  for these resonances, and  $v_\|^{res} =v_{A*}/3$ when $|k_{res}| =1.35$ for TAEs and   $|\bar{k}_{res}| = 6.15$ for HAE$_{11}$ modes at $\iota_* =0.9$. 
	We assume that $\beta_i =6.5\%$ in the plasma core, which agrees with the parameters of the Helias reactor shown in reference  \cite{Wobig}. Then  $u=1.3$ and $Q(u) =1.5$.  Now, using equation (\ref{loc1}) we obtain $|\gamma_d^{loc}|/\omega  =0.02$ for TAE modes and  $|\gamma_d^{loc}|/\omega  = 3.7\times 10^{-3}$ for HAE$_{11}$ modes. 
	
	Note that because $N\gg 1$, the frequency of MAE modes only slightly exceeds and HAE$_{11}$ frequency.  Therefore,  resonance velocity $v_\|^{res}$ produced by the $\mu /\nu =1/1$ resonance in the case of MAE modes relatively weakly differs from $v_A /3$, namely, it is   $v_\|^{res}=v_A/2.6$ for $\iota =0.9$. Therefore, the damping increases at high beta. In the Helias it can be evaluated as  $|\gamma_d^{loc}| /\omega = 2.4 \times 10^{-3}$.

\section{Summary and conclusions}
\label{sec:sum}
The results of the work can be summarized as follows.

General relations for the growth / damping rate associated with Landau mechanism  are derived.
These relations generalize  the known ones \cite{AI,LHD} by taking into account kinetic effects
in the bulk plasma and the compressibility. The latter is important for the existence of the IM modes
but plays a minor role in their damping, as shown in this work.

It is found that Landau damping of Alfv\'enic modes in stellarators plays an important role. At low $\beta$ [defined by equation (\ref{beta})], it represents a strong stabilizing mechanism of the TAE and IM modes. At high beta, which is expected in a Helias reactor \cite{Wobig}, the ion damping is  rather large not only for TAEs but although  for HAE modes and MAE modes.  The enhanced damping  is a consequence of the lack of the axial symmetry in stellarators, which leads to the resonances associated with helical harmonics of the magnetic field ($\epsilon_h \equiv \epsilon_{\mu\nu}$ with $\mu\neq 0$, $\nu\neq 0$). 
The exception is the TAE damping in a high-$\beta$ plasma, which is due to a tokamak sideband resonance.

Strong  influence of non-axisymmetric  resonances on  TAE and IM modes in the low-$\beta$ case is explained as follows. The damping rate is proportional to the square of the helical Fourier harmonics of the magnetic field, which belong to the  largest harmonics in stellarators
(in W7-X the largest helical harmonic is $\epsilon_{11}$). In contrast to this, the instability drive is proportional to the square of the toroidal harmonic, $\gamma_\alpha \propto \epsilon^2_t \ll  \epsilon^2_h$.  In addition,  the helicity-induced resonances  provide the interaction of  TAE and IM modes with a great number of particles, by  involving  the bulk plasma thermal ions, provided that $\beta_i$ satisfies a certain requirements. 

 Damping of high frequency gap modes in low-$\beta$ plasmas is realized through tokamak sideband resonance. However, its role in high frequency instabilities is rather small. First, the drive of these instabilities is stronger, $\gamma_\alpha \propto \epsilon^2_h$, whereas their damping  $\gamma_d \propto \epsilon^2_t$.  In addition,   a relatively small number of the resonant particles (electrons) is ivolved. Therefore,  high frequency gap modes in low-$\beta$ plasmas can be destabilized more easily, unless other damping mechanisms dominate.

 When $\beta$ is higher, the ion damping of MAE modes can be rather large due to the HAE resonance. 

A remarkable feature of  non-axisymmetric  resonances is that they lead to the same characteristic resonance velocities as those caused by the tokamak 
sideband resonance for TAE modes ($v_\|^{res} = v_A$ and $v_A/3$) when $\mu=\mu_0$ and $\nu =\nu_0$ \cite{AI}.  The resonance velocity 
$v_\|^{res} = v_A/3$ is connected to the plasma ion  pressure by the relation $\beta_i =1/(9u_i^2)$. It follows from here that the
 resonance velocity exactly equals  the ion thermal velocity   ($u_i=1$) when $\beta_i =1/9$, which can hardly take place in stellarators.
According to \cite{Wobig},   $\beta$ at the magnetic axis in a Helias reactor  does not exceed $13.55\%$. Assuming that $\beta_i =0.5 \beta$, we can take 
$\beta_i=6.5\%$. Then we obtain $u_i =1.3$, for which $Q$ is rather close to $Q(u_i=1)$, see figure \ref{Fig:Q}.  This explains why the damping of 
 core-localized gap modes, including high frequency modes, can be considerable in the Helias reactor.

A detailed analysis was carried out for the IM modes and TAE modes in the planned first NBI experiments on W7-X. It was found that the damping may exceed the  drive of the IM modes caused by the NBI passing ions. Therefore, it may prevent the destabilization of these modes. In order to make a more definite conclusion, the calculation of the growth rate should be carried out with a more realistic distribution function of NBI ions, and a contribution  of trapped energetic ions should be taken into account.  

It is found that our theory is consistent with an LHD experiment where two TAE modes were observed:  In this experiment 
the plasma $\beta$ was low and  the beam $\beta$ was  high, and there was
a large pressure gradient of the beam particles in the region where TAEs were located. This explained why the damping rate was not sufficient to stabilize the instability.

The conclusions drawn in this work are based on the consideration of passing particles only. The role of trapped particles is not investigated yet. One can expect that the effect of trapped electrons on the damping of high frequency modes will be not negligible. On the other hand, the instability drive may be enhanced by the   trapped energetic  ions.

\section*{Acknowledgments}
The authors thank P. Helander for a valuable discussion and A. K\"onies for providing the structure of a TAE mode in W7-X.

This work was carried out within the framework of the
EUROfusion Consortium and received funding from the
EURATOM research and training programme 2014--2018
under grant agreement No.~633053. The views and opinions
expressed herein do not necessarily reflect those of the
European Commission.
The work was also supported by Project No. PL15/17 of NASU and STCU Project No. 6392.

\end{document}